\documentclass[a4paper,twocolumn,11pt,accepted=2025-03-24]{quantumarticle}
\pdfoutput=1
\usepackage[utf8]{inputenc}
\usepackage[english]{babel}
\usepackage{subfiles}
\usepackage{multirow}
\usepackage{amssymb,amsmath,amsfonts}
\usepackage{diagbox}
\usepackage{ytableau}
\usepackage[]{hyperref}
\usepackage[mathscr]{eucal}
\usepackage[normalem]{ulem}
\usepackage{xcolor}
\usepackage{cite}

\DeclareMathOperator{\tr}{tr}

\newcommand{\ket}[1]{\left|{#1}\right\rangle}

\newcommand{\braket}[2]{\langle{#1}|{#2}\rangle}
\newcommand{\ketbrad}[1]{\left|{#1}\rangle\langle{#1}\right|}

\newcommand{\floor}[1]{\lfloor{#1}\rfloor}

\newcommand{\s}{\alpha}
\newcommand{\ihprav}{x}

\begin{document}
\title{Quantum Edge Detection}
\author{Santiago Llorens}
\email{santiago.llorens@uab.cat}
\affiliation{Física Teòrica: Informació i Fenòmens Quàntics, Departament de Física, Universitat Autònoma de Barcelona (UAB), 08193 Bellaterra (Barcelona), Spain}
\orcid{0000-0002-8643-447X}

\author{Walther González}
\email{wl.gonzalez@uniandes.edu.co}
\affiliation{Física Teòrica: Informació i Fenòmens Quàntics, Departament de Física, Universitat Autònoma de Barcelona (UAB), 08193 Bellaterra (Barcelona), Spain}
\affiliation{Grupo de Óptica Cuántica, Departamento de Física, Universidad de los Andes, Cra. 1, 18A-12, Bogotá, Colombia}

\author{Gael Sent\'is}
\email{gael.sentis@uab.cat}
\affiliation{Física Teòrica: Informació i Fenòmens Quàntics, Departament de Física, Universitat Autònoma de Barcelona  (UAB), 08193 Bellaterra (Barcelona), Spain}
\affiliation{Ideaded, Carrer de la Tecnologia, 35, 08840 Viladecans, Barcelona, Spain}
\orcid{0000-0002-4982-6570}

\author{John Calsamiglia}
\email{john.calsamiglia@uab.cat}
\affiliation{Física Teòrica: Informació i Fenòmens Quàntics, Departament de Física, Universitat Autònoma de Barcelona (UAB), 08193 Bellaterra (Barcelona), Spain}
\orcid{0000-0003-1735-1360}

\author{Ramon Muñoz-Tapia}
\email{ramon.munoz@uab.cat}
\affiliation{Física Teòrica: Informació i Fenòmens Quàntics, Departament de Física, Universitat Autònoma de Barcelona (UAB), 08193 Bellaterra (Barcelona), Spain}
\orcid{0000-0002-3048-9236}

\author{Emili Bagan}
\email{emili.bagan@uab.cat}
\affiliation{Física Teòrica: Informació i Fenòmens Quàntics, Departament de Física, Universitat Autònoma de Barcelona (UAB), 08193 Bellaterra (Barcelona), Spain}
\orcid{0000-0002-7900-3567}

\maketitle

\begin{abstract}
{\boldmath
This paper introduces quantum edge detection, aimed at locating boundaries of quantum domains where all particles share the same pure state. Focusing on the 1D scenario of a string of particles, we develop an optimal protocol for quantum edge detection, efficiently computing its success probability through Schur-Weyl duality and semidefinite programming techniques. We analyze the behavior of the success probability as a function of the string length and local dimension, with emphasis in the limit of long strings. We present a protocol based on square root measurement, which proves asymptotically optimal. Additionally, we explore a mixed quantum change point detection scenario where the state of particles transitions from known to unknown, which may find practical applications in detecting malfunctions in quantum devices.}
\end{abstract}
\maketitle
\section{Introduction}

In recent years, the rapid advancement of quantum technologies has opened up new avenues for exploring the unique properties and potential applications of quantum systems. 
These advancements span various domains, including~quantum computation~\cite{arute_quantum_2019} and simulation~\cite{georgescu_quantum_2014}, \mbox{quantum} communication~\cite{khatri_principles_2024}, quantum sensing~\cite{degen_quantum_2017} and metrology~\cite{giovannetti_quantum-enhanced_2004}, quantum machine learning~\cite{dunjko_machine_2018,peral-garcia_systematic_2024}, and eventually, the realization of a quantum internet~\cite{kimble_quantum_2008}, which would enable not just secure key distribution~\cite{gisin_quantum_2002} but also distributed quantum computing. While these advancements present exciting opportunities, they also introduce challenges that cannot be adequately addressed by classical means, necessitating specialized quantum methodologies for effective resolution (see, e.g., \cite{mcclean_barren_2018} and \cite{preskill_quantum_2018}). The precise classification of quantum data and the detection of abrupt changes in this data, which can signal malfunctions or anomalies requiring immediate attention, are important examples. These protocols also serve as primitives in quantum data processing and find application across various domains.

Classifying quantum  
data is paramount for future developments such as the quantum internet. To achieve this, automated tools are necessary to make the classification process practical. Traditional classical and quantum techniques often fall short in handling the complexity of this task. Quantum machine learning emerges as a promising solution to these challenges~\cite{monras_inductive_2017,fanizza_optimal_2019,wall_tensor-network_2022,zhu_flexible_2022}, offering a pathway to automate and streamline the classification process. Similarly, the detection of sudden changes in quantum systems~\cite{sentis_quantum_2016,sentis_exact_2017} should be performed in an automated manner to eliminate the need for human monitoring. Once again, quantum machine learning techniques could be valuable tools in this regard~\cite{liu_quantum_2018}. 
By exploiting these techniques, potential issues can be proactively addressed, and the integrity of quantum data can be maintained in various applications.

In a recent paper~\cite{sentis_unsupervised_2019}, the concept of unsupervised classification of quantum data was introduced, focusing on systems with unknown quantum states. The paper presents an optimal single-shot protocol for binary classification, capable of automatically sorting disordered input arrays with minimal assumptions while preserving partial information. It is demonstrated that this protocol surpasses classical methods for dimensionality three or higher. Furthermore, while the quantum data classification protocol runs efficiently on a quantum computer, the equivalent task with classical data involves solving an NP-complete problem. Its potential applications in high-energy physics are discussed in~\cite{guan_quantum_2021}.

In a similar vein, in this paper, we introduce a new concept that we term `quantum edge detection'. It is a task whereby the boundaries of quantum domains, where all particles share the same pure state, are precisely located. This addresses the challenge of uncovering the spatial layout of complex quantum systems, such as cold ions trapped in a lattice or spin-lattices subject to multiple localized magnetic fields. Specifically, here we focus on the simplest version of this problem in 1D, where we have an array of qudits arranged in two domains, and we derive the ultimate limits imposed by nature on how successfully this task can be accomplished. We leave the considerably more demanding 2D and 3D problems for further work.

Our approach benefits from recent advancements in genuinely quantum technology, including techniques like unsupervised classification, briefly discussed above, where now some information about the structure of the states is provided. These advancements enable us to discern the spatial configuration of the quantum system, akin to how classical edge detection~\cite{canny_computational_1986} or binary image segmentation~\cite{pal_review_1993,minaee_image_2022} is employed in computer vision. 
In some sense, quantum edge detection can be viewed as an extension of these concepts when the pixels that make up the image are quantum. This
differs fundamentally from recent methods for edge detection in classical image processing that use quantum algorithms~\cite{PhysRevX.7.031041}.

We consider a string of $N$ qubits comprising two distinct domains, each composed of particles in identical pure states. We assume that~the observer is unaware of the specific states; the~only known information is their uniformity within each region. Our task  
is to identify the position of the boundary between the two domains. This 1D version of quantum edge detection bears some similarities with the Quantum Change Point (QCP) problem~\cite{sentis_quantum_2016}. Leaving aside that the focus of the QCP detection protocol is on time series rather than on spatial arrangements of particles, the main difference is that the states prior and after the change takes place were assumed to be known.

Any identification procedure we may envision necessarily involves performing quantum measurements on the string, and it will fail with some probability. This is because the states corresponding to the various possible locations of the boundary or edge will usually not be orthogonal and, hence, not perfectly distinguishable. 

We choose to evaluate the procedure's performance based on the success probability of correctly identifying the edge. 
Since we lack knowledge about the states of the particles in each domain beyond their purity, the procedure we seek must be universal~\cite{fanizza_universal_2022}, meaning it must be independent of these quantum states. Its effectiveness will vary depending on the specific states of the domains in the given string. Therefore, we can assess the overall performance of the procedure by either averaging the success probability across all possible cases (states) or considering the success probability in the worst case. In our scenario, however, the worst case occurs when the states in the two domains are identical (no edge to detect), resulting in a trivial figure of merit.

We define the optimal edge detection protocol as one that maximizes the average success probability, assuming a uniform prior probability distribution for the location of the edge along the string.
The ultimate limits mentioned above will be given by this quantity, as a function of $N$ and the local dimension $d$.
It can be argued, following the reasoning in~\cite{sentis_unsupervised_2019}, that the optimality of the protocol will also extend to a broad family of figures of merit, which include the Hamming distance and the mean square error.

Let us summarize our main results.

We first uncover the (highly-symmetric) structure of quantum edge detection, enabling us to exploit the Schur-Weyl duality and streamline the problem to pure-state discrimination. This involves performing a global measurement on the entire string, projecting its state onto the ${\rm SU}(d)$ invariant subspaces. The resulting conditioned states are pure and in one-to-one correspondence with the location of the edge. By optimally discriminating these states, we achieve optimal edge detection.

This optimization can be formulated as an $N$-dimensional Semidefinite Program (SDP) problem~\cite{vandenberghe_semidefinite_1996} and executed efficiently, despite the exponential growth of the dimension of the conditioned states with $N$. The approach involves using the Gram matrix, which has size $N^2$, in a manner akin to the kernel trick in machine learning~\cite{bishop_pattern_2006}, allowing for efficient execution.

Next, we derive an asymptotic expression for the averaged success probability of the optimal protocol, valid in the limit of large $N$. It is seen to approach a finite value in this limit. Furthermore, this value agrees with the optimal success probability of detection, assuming that the states of the domains are known and averaged over all possible pure states. This average can be straightforwardly computed from the results on QCP in~\cite{sentis_quantum_2016}.

The dependence of the optimal success probability on $N$ is quite revealing. One might expect that increasing the length of the string makes edge detection harder, since the number of possible locations of the edge increases. However, we observe that for sufficiently large strings, the success probability is actually an increasing function of $N$. This behavior can be intuitively understood by noticing that larger domains potentially provide more information about the state of the particles they consist of, making edge detection easier. This tradeoff results in an increasing success probability. This is in contrast with the QCP, for which the success probability decreases with $N$. Simulations validate this intuition: We have compelling numerical evidence that a two-step strategy, involving the estimation of the known state of each domain from a vanishing fraction of particles (e.g., of the order of~$\sqrt N$\,), followed by the application of the optimal (non-local) QCP protocol for edge localization, is asymptotically optimal.

Lastly, we consider an intermediate scenario where the state of the particles in one of the domains is known to the observer, while that of the particles in the other domain is unknown. This scenario is better understood as a QCP problem, where a source initially produces a string of particles with fully specified states but undergoes a failure at some point, resulting in a change of states and loss of specification. Such scenarios could find practical application in detecting malfunctions in devices designed to deliver batches of particles in identical states.
The success probability shows similar behavior and, as expected, falls between those of the two extreme cases previously considered, where the states in the domains are either both known or both unknown, thus converging to the same limit asymptotically.

This paper is organized as follows. Section~\ref{sec:structure} presents a detailed mathematical description of the problem. The derivation of the optimal protocol and an assessment of its performance are discussed in Section~\ref{sec:methodology}. In Section~\ref{sec:knownunknown}, we shift our focus to the alternative scenario where the state of the particles in one of the domains is known. The paper concludes with final remarks and an outlook for future investigations.
Additional technical details can be found in Appendices A--C.

\section{Structure of the problem}
\label{sec:structure}

Let $\ket{\phi_0},\ \ket{\phi_1}\in{\mathbb C}^d$ denote the states of the qudits in each domain and $k=1,2,\ldots, N$ the position of the edge we aim to detect. Under the assumptions discussed in the introduction, the global state of the string can be expressed as:
\begin{equation}
\label{eq:ordered sequence}
\ket{\Phi_k} = \underbrace{\ket{\phi_0}\otimes\dots\otimes\ket{\phi_0}}_{N-k} \otimes \overbrace{\ket{\phi_1}\otimes\dots\otimes\ket{\phi_1}}^{k}.
\end{equation}

The detection protocol, including post-pro\-cess\-ing, is defined by a Positive Operator-Valued Measure (POVM), whose elements, ${E_k}$, are associated with the outcomes of a measurement, indicating the possible locations of the edge along the string. To find the optimal protocol, as discussed in the introduction, we aim to maximize the average success probability
\begin{equation}
P_{\rm s} ={1\over N}\int {\rm d}\phi_0\, {\rm d}\phi_1\sum_k\tr\left(\ketbrad{\Phi_k}E_k\right).
\label{P_s def}
\end{equation}
 As explained, this approach considers a uniform prior distribution of the edge between the two domains, reflecting a total lack of information about its location. Additionally, the average over the states $\ket{\phi_0}$ and $\ket{\phi_1}$ acknowledges that they are unknown to the user.

The universality of the detection protocol is ensured by the independence of the positive operators ${E_k}$ from the states of the particles in the string. This independence allows ${E_k}$ to be moved outside the integral in~(\ref{P_s def}), thereby enabling the definition of effective states $\rho_k$ for the string. Specifically,
\begin{multline}
\rho_k= \int \left(|\phi_0\rangle\langle\phi_0|\right)^{\otimes(N-k)}\otimes\\
\left(|\phi_1\rangle\langle\phi_1|\right)^{\otimes k} \mathrm{d}\phi_0\, \mathrm{d}\phi_1 .
\label{eq:hypothesisrho}
\end{multline}
A straightforward application of the Schur lemma yields 
\begin{align}
\rho_k &= {\openone^{\text{sym}}_{N-k}\otimes \openone^{\text{sym}}_{k}\over  d_{N-k}^{\,\mathrm{sym}}\,d_{k}^{\,\mathrm{sym}}  },
\label{eq:twosymmetric}
\end{align}
where $\openone^{\text{sym}}_n$ is the projector onto the symmetric subspace of $n$ qudits, whose dimension is given~by
\begin{equation}
\label{eq:dim-sym}
    d_n^{\,\mathrm{sym}}=\binom{d+n-1}{d-1} \,.
\end{equation}
In the Schur basis~\cite{fanizza_universal_2022,sentis_unsupervised_2019}, the states $\rho_k$ take on a convenient block-diagonal form. This structure results from the identity
   \begin{align}
  \openone^{\text{sym}}_{N-k}\otimes \openone^{\text{sym}}_{k}&= \bigoplus_{\boldsymbol{\lambda}} \openone_{\boldsymbol{\lambda}}\otimes \Omega^{\boldsymbol{\lambda}}_k ,
    \label{eq:schurbasis}
\end{align}
which, in turn, is a direct consequence of the Schur-Weyl duality. Here, $\boldsymbol{\lambda}$ labels the ir\-re\-duc\-ible representations (irreps) arising from the joint action of the unitary group $\textrm{SU}(d)$ and the symmetric group $S_N$ on the vector space $(\mathbb{C}^d)^{\otimes N}$ of the entire string, where the symmetric group operates by permuting the particles within the string. Conveniently, $\boldsymbol{\lambda}$ is usually identified with partitions of $N$, or equivalently with Young Diagrams (YD) of~$N$ boxes.
In our case, only~two-row YD arise because, as shown in Eq.~(\ref{eq:twosymmetric}), the density matrices~$\rho_k$ act on the tensor product of two symmetric  irreducible subspaces (each represented by one-row YD)~\cite{sentis_unsupervised_2019}. 
Consequently, all required labels have the simple form $\boldsymbol{\lambda}=[N-\lambda,\lambda]$ with $\lambda=0,1,2,\dots, \floor{N/2}$, where $\floor{\cdot}$ is the floor function. Hereafter, we will write the label $\boldsymbol{\lambda}$ simply as $\lambda$. Each projector~$\Omega_{k}^{\lambda}$ acts on the $S_N$ irrep subspace $\lambda$ and is rank-1, which again follows from the form of $\rho_k$ shown in Eq.~(\ref{eq:twosymmetric}), thus, it can be written as $\Omega^{\lambda}_k=|\Omega^{\lambda}_k\rangle\langle\Omega^{\lambda}_k|$. Note that the state $\rho_k$ has support in irreps with $\lambda=0,1,2,\dots, \min\{N-k,k\}$. 

Given the structure of $\rho_k$ as shown in Eq.~\eqref{eq:schurbasis}, one can choose, without loss of generality, POVM operators of the form
\begin{equation}
E_k = \bigoplus_{\lambda} \openone_{\lambda}\otimes \mathscr{E}^{\lambda}_k,
\end{equation}
which implies that it suffices to consider two-step protocols starting with a projective measure\-ment~onto the irrep subspaces. This mea\-surement, referred to as weak Schur sampling, is represented by the projectors $\{\openone_{\lambda} \otimes \openone^{\lambda}\}_{\lambda=0}^{\lfloor N/2 \rfloor}$ and outputs a value,~$\lambda^*$, for the irrep label $\lambda$. Subsequently, a measurement with POVM elements $\{\openone_{\lambda^*}\otimes\mathscr{E}^{\lambda^*}_k\}_{k\in K_{\lambda^*}}$ is executed on the posterior state belonging to the~$\lambda^*$ subspace. For a generic value of $\lambda$, $K_\lambda$ is the range of values of $k$ for which $\rho_k$ has support in the $\lambda$ subspace, $\lambda\le k\le N-\lambda$.

In this second step of the detection protocol, we effectively implement a discrimination strategy among the set of pure states $\{|\Omega_k^\lambda\rangle\}_{k\in K_\lambda}$ with a prior probability mass function given by $\eta^\lambda_k/\sum_{k'} \eta^\lambda_{k'}$.
In this equation, $\eta_k^\lambda$ represents the joint probability of the edge being located at $k$ and obtaining the outcome $\lambda$ from the weak Schur sampling. This probability is expressed as: 
\begin{equation}
    \label{eq:priors}
    \eta^{\lambda}_{k}=
    \begin{cases}
    \dfrac{s_\lambda}{N  d_{N-k}^{\,\mathrm{sym}}\,d_{k}^{\,\mathrm{sym}} }, & k\in K_\lambda\\[1em]
    0, & \mbox{otherwise,}
    \end{cases}
\end{equation}
where $s_\lambda$ denotes the dimension of the irrep $\lambda$, namely $s_\lambda=\text{tr}(\openone_\lambda)$, explicitly given by the expression:
\begin{multline}
    s_\lambda=\frac{N-2\lambda+1}{N-\lambda+1}\binom{d+\lambda-2}{d-2}\times\\
    \binom{d+N-\lambda-1}{d-1} \,.
\end{multline}
It can be observed that $d^{\,\text{sym}}_n$ is equal to $s_0$ evaluated at $N=n$, as expected.

Combining the above results, the average success probability, which we aim to maximize, reads
\begin{equation}
P_{\rm s}=\sum_{\lambda=0}^{\floor{N/2}} \sum_{k\in K_\lambda} \eta^\lambda_k  \tr\left(\Omega^\lambda_k{\mathscr E}^\lambda_k\right).
\end{equation}
Although algorithms exist that could theoretically compute $|\Omega^\lambda_k\rangle$, the dimension of the irrep spaces to which these states belong grows exponentially with $N$~\cite{olaya_kronecker_2024}, making optimization over $\{{\mathscr E}^\lambda_k\}_{k,\lambda}$ via direct computation infeasible even for small values of $N$.

To address this challenge, we will use the Gram matrix ${\sf G}^\lambda$ of the set of pure states $\{|\Omega^\lambda_k\rangle\}_{k\in K_\lambda}$ for each value of $\lambda$. This matrix contains all the information necessary for optimizing the discrimination task. Its entries are given by
\begin{equation}
\label{eq:G-tilde}
     G_{k k'}^\lambda  =\sqrt{\eta^{\lambda}_{k}\eta^{\lambda}_{k'}}\ \langle\Omega^{\lambda}_k|\Omega^{\lambda}_{k'}\rangle.
 \end{equation}
Since ${\sf G}^\lambda$ is an $|K_\lambda| \times |K_\lambda|$ matrix, the computational complexity of the optimization problem becomes polynomial in $N$.
This approach is reminiscent of the kernel trick often employed in machine learning, where the highly dimensional states $|\Omega^\lambda_k\rangle$ play the role of feature vectors.

A closed-form expression of the Gram matrix is derived in Appendix \ref{app:sixyeys}. It takes the simple form:
\begin{equation}
\label{eq:overlap-u-u}
G^\lambda_{kk'} = v^\lambda_k u^\lambda_{k'}, \quad k\leq k'
\end{equation}
(for $k' \leq k$, exchange $k$ and $k'$), where:
\begin{equation}
\label{eq:def u v}
u^\lambda_{k} = \sqrt{\eta^\lambda_k\,\dfrac{\binom{N-k}{\lambda}}{\binom{k}{\lambda}}}, \quad
v^\lambda_k = \sqrt{\eta^\lambda_k\,\dfrac{\binom{k}{\lambda}}{\binom{N-k}{\lambda}}}.
\end{equation}
Note that the Gram matrix ${\sf G}^\lambda$ can be represented as ${\sf G}^\lambda = \operatorname{triu}({\sf u}{\sf v}^T) + \operatorname{tril}({\sf v}{\sf u}^T)$, where $\operatorname{tril}$ ($\operatorname{triu}$) denotes lower (upper, including the diagonal) triangular part. Here, $\sf u$ and $\sf v$ are column vectors with components given by~(\ref{eq:def u v}). This property ensures that ${\sf G}^\lambda$  is semiseparable~\cite{mastronardi_matrix_2008}, offering significant computational advantages that complement the complexity reduction achieved by using the Gram matrix formulation. In particular, we will exploit the fact that the inverse of a semiseparable matrix is tridiagonal, resulting in sparsity.
Appendix~\ref{app:sixyeys} provides a proof of the semiseparability of ${\sf G}^\lambda$ using group theoretical arguments.

In the following section, we compute the maximum average success probability $P_{\rm s}$ via numerical optimization, using SDP, for values of $N$ up to~$55$. We then compare these results with those obtained by employing the square root (or pretty good) measurement~\cite{hausladen_pretty_1994,hausladen_classical_1996} to discriminate among the states $|\Omega^\lambda_k\rangle$. These values offer a remarkably good lower bound for $P_{\rm s}$, which we can evaluate for much larger $N$ values. We subsequently draw conclusions regarding the optimality of the latter strategy.

\section{Methodology and numerical results}
\label{sec:methodology}
\subsection{Numerical results}

The inherent symmetries of 1D quantum edge detection enable us to simplify our optimization task into a series of more manageable pure-state discrimination sub-problems, each framed within an irrep subspace $\lambda$. This streamlined approach significantly reduces the complexity involved.

In general, quantum state discrimination problems lack known analytical solutions, except for two states~\cite{helstrom_quantum_1969} or for highly symmetric sets of states~\cite{barnett_minimum-error_2001,chiribella_covariant_2004,sentis_online_2022}. Since our sub-problems do not fit into any of these simple cases, we must resort to numerical optimization.

Without relying on Schur-Weyl duality, a direct numerical method would encounter exponential complexity due to the high dimensionality ($d^N$) of the effective mixed states $\rho_k$. However, Schur-Weyl duality alone cannot sufficiently tame the problem without invoking the Gram matrix formulation. As discussed in Sec~\ref{sec:structure}, using~${\sf G}^\lambda$, with a size at most $N^2$, provides an exponential advantage, enabling us to efficiently employ SDP optimization, whose complexity scales only polynomially with the size of ${\sf G}^\lambda$.

The trick involves viewing the $N$ columns of the symmetric and positive definite matrix $\sqrt{{\sf G}^\lambda}$, each denoted as $\raisebox{.2em}{$[$}\sqrt{{\sf G}^\lambda}\,\raisebox{.2em}{$]_k$}$, as representations of the $N$ unnormalized states $\sqrt{\eta^\lambda_k}|\Omega^\lambda_k\rangle$. The consistency of this representation can be verified by observing that
\begin{equation}
\raisebox{.2em}{$[$}\sqrt{{\sf G}^\lambda}\,\raisebox{.2em}{$]^T_k$}\raisebox{.2em}{$[$}\sqrt{{\sf G}^\lambda}\,\raisebox{.2em}{$]_{k'}$}=\raisebox{.2em}{$\big[($}\sqrt{{\sf G}^\lambda}\,\raisebox{.2em}{$)^2\big]_{kk'}$} %
=G^\lambda_{kk'},
\end{equation}
where the superscript $T$ denotes transpose.
The success probability of the optimal discrimination strategy on a given irrep $\lambda$ can then be computed~as
\begin{gather}
P_\lambda \!=\max_{\{{\sf E_k}\}}\sum_{k\in K_\lambda}\!\! 
\raisebox{.2em}{$[$}\sqrt{{\sf G}^\lambda}\,\raisebox{.2em}{$]^T_k$}%
{\sf E}_k\, \raisebox{.2em}{$[$}\sqrt{{\sf G}^\lambda}\,\raisebox{.2em}{$]_k$} \nonumber
\\[-.1em]
\mbox{subject to}   \label{eq:SDP lambda}
\\[.2em]
 {\sf E_k}\ge 0,\quad  \sum_{k\in K_\lambda}{\sf E_k}=\openone^\lambda ,
 \nonumber
\end{gather}
where ${\sf E}_k$ is a $|K_\lambda|\times|K_\lambda|$ matrix representation of the POVM operator ${\mathscr E}^\lambda_k$.
We observe that the optimization problem in Eq.~(\ref{eq:SDP lambda}) takes the form of a SDP. The average success probability of the optimal detection protocol is then given by
\begin{equation}
\label{eq:SDP total}
P_{\rm s}=\sum_{\lambda=0}^{\floor{N/2}}P_\lambda .
\end{equation}
Here $P_\lambda$ denotes the (joint) probability of successfully detecting the edge and obtaining the outcome $\lambda$ from the weak Schur sampling.

\begin{figure}[htb]
\centering
\includegraphics[width=.97\columnwidth]{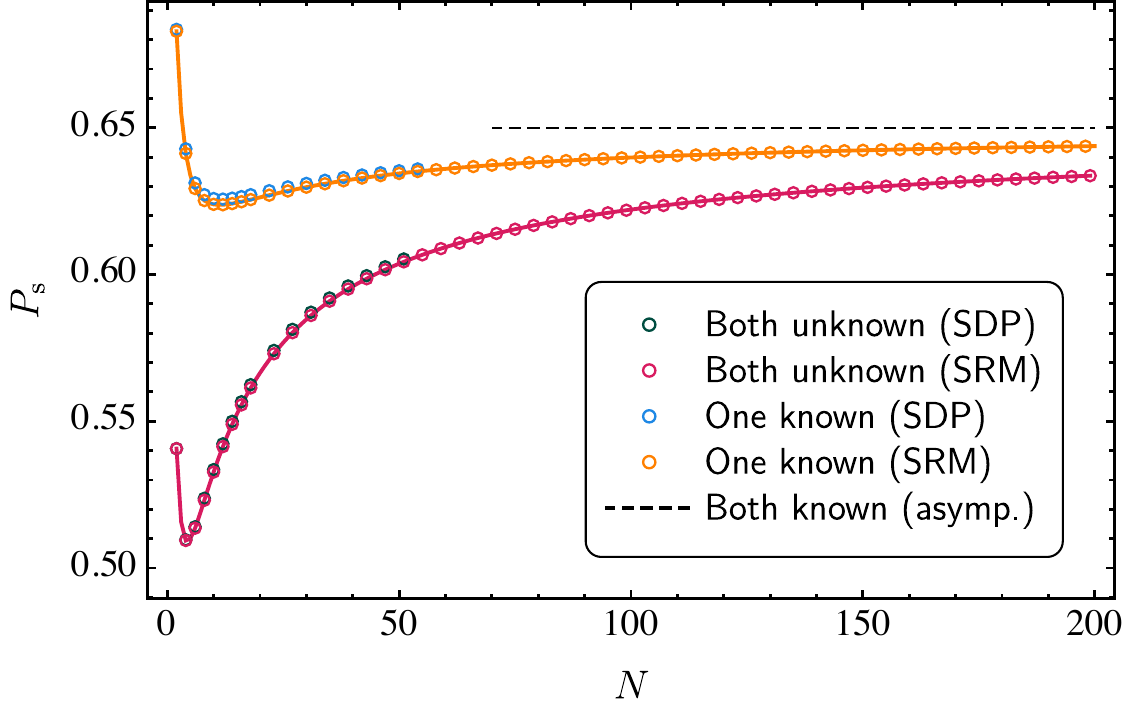}
\caption{Success probability as a function of the string length $N$ for the scenarios considered in this paper: when all states are unknown (green and pink circles) and when the states of one domain are known (blue and orange circles). The maximum success probability data points, obtained in both scenarios through SDP optimization, are shown in green and blue, respectively, while the SRM success probability points are in pink and orange, respectively. The black dashed line represents the average asymptotic success probability derived from the QCP, assuming knowledge of all states. All data points correspond to $N$ ranging from $2$ to $18$ in steps of $2$, and from $22$ to $198$ in steps of $4$.}
 \label{fig:all_data_ylf}
\end{figure}

The green points in Fig. \ref{fig:all_data_ylf} illustrate the results of our calculation, obtained using Eqs.~(\ref{eq:SDP lambda}) and~(\ref{eq:SDP total}) for $d=2$ (qubits). Similar trends are observed for other values of $d$. These points represent the ultimate limit to the achievable success probability in detecting the edge of a domain along a string of quantum particles.

While not always optimal~\cite{dalla_pozza_optimality_2015}, the Square Root Measurement (SRM)~\cite{hausladen_pretty_1994,hausladen_classical_1996} often yields accurate lower bounds on success probability across numerous discrimination problems. It has been shown to offer an asymptotically optimal quantum change point detection protocol in~\cite{sentis_quantum_2016}. That work also presents general conditions for the SRM's optimality and provides bounds on the success probability it achieves. However, in the present case, the SRM does not meet the required conditions for these results to hold. In spite of this, we will verify its asymptotic optimality for edge detection.

The POVM elements of the SRM are defined as
\begin{equation}
\mathscr{E}^\lambda_k = \Lambda_\lambda^{-1/2}\left(\eta^\lambda_k\,\Omega^\lambda_k\right)\Lambda_\lambda^{-1/2},
\end{equation}
where $\Lambda_\lambda = \sum_k \eta^\lambda_k\,\Omega^\lambda_k$. In our lower-dimensional representation, this fixes the definition of the matrix elements of~$\mathsf{E}_k$ in Eq.~(\ref{eq:SDP lambda}) to $[{\sf E}_k]_{ii'} = \delta_{ki}\delta_{ki'}$ independently of $\lambda$. With this choice, the expression for $P^{\rm SRM}_\lambda$ simplifies to
\begin{equation}
\label{eq: SRM lambda}
P^{\rm SRM}_\lambda=\sum_{k\in K_\lambda}\!\!  \raisebox{.1em}{$\Big($}\raisebox{.2em}{$[$}\sqrt{{\sf G}^\lambda}\,\raisebox{.2em}{$]_{kk}$}\raisebox{.1em}{$\Big)^2$}.
\end{equation}
In simple terms, the joint probability of obtaining~$\lambda$ and succeed amounts to the sum of the squares of the diagonal entries of $\sqrt{{\sf G}^\lambda}$. The success probability $P^{\rm SRM}=\sum_\lambda P^{\rm SRM}_\lambda$ gives a lower bound to $P_{\rm s}$ since it results from a particular choice for~${\sf E}_k$ in Eq.~(\ref{eq:SDP lambda}).

Since no optimization is required to compute~$P^{\rm SRM}_\lambda$, it becomes feasible to evaluate the success probability $P^{\rm SRM}$ for significantly larger values of $N$. Additionally, the SRM equips us with the tools to derive analytical results for the asymptotic behavior of $P_{\rm s}$ as $N$ grows large, a topic we discuss in the next subsection.

In Fig.~\ref{fig:all_data_ylf}, the pink points, connected by lines, depict the results of our numerical evaluation of $P^{\rm SRM}$ for $d=2$ %
(similar patterns hold for other values of $d$). It is noteworthy~that the pink points  differ almost imperceptibly from~the green ones, which represent the optimal protocol, indicating that the SRM is nearly optimal.
Both sets of points fall underneath the dashed line but gradually approach it. This line represents a scenario where the quantum states of the domains are known and this information is used for edge detection. More precisely, it depicts the average success probability over $|\phi_0\rangle$ and $|\phi_1\rangle$ in Eq.~(\ref{P_s def}) when the measurement is allowed to depend on these states and $N\to\infty$. This average, that we denote as~$P_{\rm s}^{\rm known}$, can be easily computed (see below) using the results from the QCP derived in \cite{sentis_quantum_2016} and also serves as an upper bound to $P_{\rm s}$ in the scenario considered in this paper. The approaching of the pink points to this line also showcases the asymptotic optimality of the~SRM.

Intuition might suggest that the success probability should monotonically decrease with $N$, as one could argue that with a longer string of particles, there would be more possible locations for the edge, making the detection task more challenging. This is indeed the behavior of the success probability for the QCP, akin to the scenario where the states of the domains are known. However, Fig.~\ref{fig:all_data_ylf} disproves this intuition, except for very short strings of fewer than eight particles. For longer strings ($N \geq 8$), the trend is reversed: edge detection becomes more successful as the number of particles grows larger. This suggests that the detection protocol learns from~the unknown states of the domains, overcoming the increasing difficulty of discriminating possible edge locations.

To test this idea, we simulated an estimate-and-discriminate strategy where a small fraction (of order $\sqrt{N}$) of particles from each end of the string is used to estimate the states of the domains. This information is then fed into the QCP protocol to precisely locate the edge. While we can only perform this simulation for modest string lengths, the results align with those shown in Fig.~\ref{fig:all_data_ylf}, with the success probability approaching the value achieved by the optimal edge detection protocol.

\subsection{Asymptotic performance}
\label{sec:optimal}
In this section, we assess the performance of the optimal detection protocol in the asymptotic regime of large $N$ by computing the limiting value of the success probability. The calculation is performed indirectly, by verifying that the lower bound $P^{\rm SRM}_s$ provided by the SRM strategy matches the upper bound $P_{\rm s}^{\rm known}$ resulting from averaging the success probability under the assumption of known domain states. This alignment, shown in Table~\ref{ebc11.06.22-t2} below, not only reveals the value of the maximum success probability but also confirms the asymptotic optimality of the SRM strategy.

To this end, we will show that the success probability has an asymptotic expansion of the form 
\begin{equation}
\label{eq: pert exp}
P^{\rm SRM}_s\sim p_0(d)+{p_1(d)\over N}+O(N^{-2})
\end{equation}
and present a method for computing the coefficients $p_n(d)$, $n=0,1,2,\dots,$~to arbitrary accuracy. The method shares similarities with the application of perturbation theory in fields like high-energy physics, particularly relying on techniques to expedite the convergence of asymptotic series through the utilization of Pad\'e approximants.

Here, we only provide an overview of the method, focusing on computing $p_0(2)$, the leading coefficient for qubits ($d=2$), with~$0.03\%$ uncertainty. We will see that the computed value coincides with the upper bound $P^{\rm known}_s$, within the specified accuracy. 
Results for larger dimensions ($d=3,4,8$) are also given, but technical details are deferred to Appendix~\ref{app:semianalytic}.

The essence of our calculations lies in a `perturbative expansion' of the inverses of the Gram matrices ${\sf G}^\lambda$, rescaled appropriately. We recall that these inverses are tridiagonal matrices. The sparsity of these matrices, with their off-diagonal entries treated as perturbations, is crucial for efficiently keeping track of the orders of the expansion.

The rescaling of the Gram matrices is performed as follows:
\begin{equation}
\label{eq:rescaling}
 \tilde{\sf G}^\lambda=\frac{(N/2)^2}{N-2\lambda+1}{\sf G}^\lambda,
\end{equation}
ensuring that $\tilde{\sf G}^\lambda=\openone+O(1/N)$ for large $N$. Thus, $(\tilde{\sf G}^\lambda)^{-1}=\openone+\Delta_\lambda$, where $\Delta_\lambda$ contains non-zero terms only in the super- and sub-diagonals, each of order $O(1/N)$. Utilizing the binomial series, we obtain
\begin{multline}
\label{eq:sqrt-expansion}
    \sqrt{\tilde{\sf G}^\lambda}=\left(\openone+\Delta_\lambda\right)^{-1/2}\\
    =\openone+\sum_{r=1}^\infty\binom{-1/2}{r}(\Delta_\lambda)^r.
\end{multline}
In practical calculations, the series is truncated to the desired order of approximation, rescaled back, and then substituted into Eq.~(\ref{eq: SRM lambda}). 
This procedure yields an asymptotic expression for the joint probability $P^{\rm SRM}_\lambda$, whose marginal probability (over $\lambda$) provides the desired asymptotic series~(\ref{eq: pert exp}) for the success probability. The initial terms can be expressed~as:
\begin{multline}
\label{eq:P lambda exp}
P^{\rm SRM}_\lambda={4(2j+1)^2 \over N^3}   \left[1-{4\over N}
+     \vphantom{ {8(j^2\!+\!j\!+\!12)\over 3N^3}}   \right. \\[.5em]
{12\over N^2}- {8(j^2+j+12)\over 3N^3}-\\[.5em]
 \left .  {4j^4+8j^3+\dots\over 15 N^4} + \dots \right].
\end{multline}
Here, $j = N/2 - \lambda$ denotes the total spin quantum number of the string of particles (qubits), which serves as an alternative and commonly used label for the irreps of ${\rm SU}(2)$.

\begin{figure}[htb]
\centering
\includegraphics[width=.95\columnwidth]{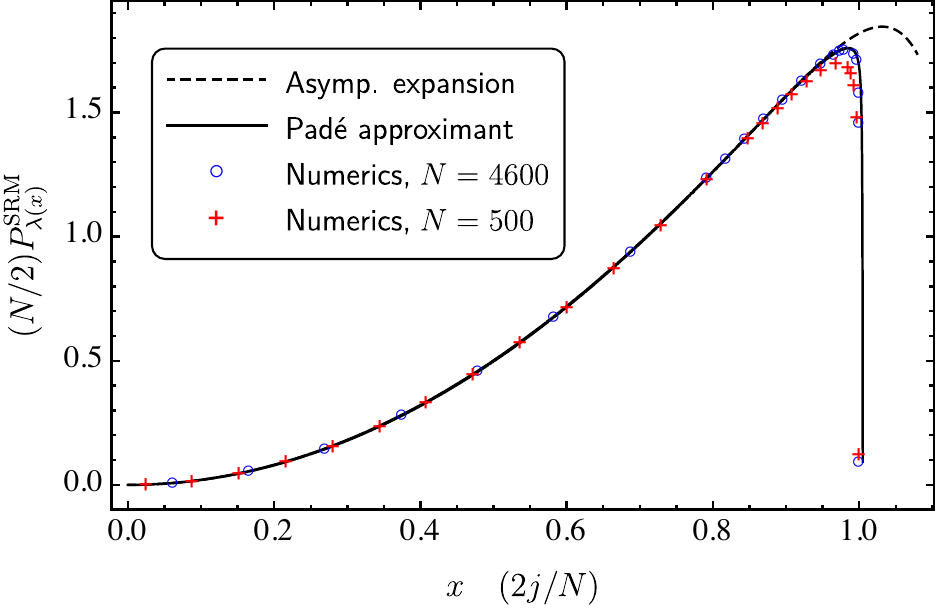}
\caption{Joint probability $P^{\rm SRM}_{\lambda(\ihprav)}$ plotted against the scaled angular momentum $\ihprav = 2j/N$. The dashed curve represents the asymptotic power series truncated at $O(\ihprav^{30})$, while the solid curve depicts its $[14/14]$ Padé approximant. These curves closely match each other, deviating only near $\ihprav=1$. The range of~$x$ extends beyond its physical domain (the unit interval) to show this deviation more clearly. The red crosses and blue circles denote numerical evaluations of the joint probability at selected points, corresponding to $N=500$ and $N=4600$ respectively.  The solid line is seen to fit the numerical points corresponding to the larger value of $N$ almost perfectly.}
 \label{fig:figEBC}
\end{figure}

The calculation of the marginal, $\sum_j P^{\rm SRM}_{\lambda(j)}$, is involved due to the scaling of the range of values of $j$ with $N$, resulting in infinitely many terms in the expansion~(\ref{eq:P lambda exp}) that contribute to the coefficient $p_0(2)$. This can be addressed by introducing a scaled version of $j$ as $\ihprav:=j/(N/2)$. In the limit $N\to\infty$, the variable $\ihprav$ can be thought of as a real variable whose range is $[0,1]$. In terms of $\ihprav$, Eq.~(\ref{eq:P lambda exp}) takes the form
\begin{equation}
{N\over2}P^{\rm SRM}_{\lambda(\ihprav)}=\sum_{r=1}^\infty a_r \ihprav^{2r} ,
\label{ebc08.06.22-3}
\end{equation}
and the leading-order term $p_0(2)$ can be computed as
\begin{equation}
\label{eq: int P_lambda}
p_0(2)=\int_0^1{N\over2}P^{\rm SRM}_{\lambda(\ihprav)}\, {\rm d}\ihprav .
\end{equation}
However, a difficulty arises as we have not identified a closed expression for the general term of the sequence $\{a_r\}_{r\in{\mathbb N}}$, and the truncated series, which we managed to compute up to $O(\ihprav^{30})$, is unable to capture the singular behavior of the integrand near $\ihprav=1$ with sufficient accuracy (dashed line in Fig.~\ref{fig:figEBC}). 
Padé approximants can be used to speed up the convergence of the series as they are known to accurately describe the behavior of asymptotic expansions near singular points~\cite{bender_advanced_1999}. 

In brief, for a given function $f(\ihprav)$, its Padé approximant of order $[n/m]$ is the rational function
\begin{equation}
{}[n/m]_f(\ihprav):=
{\sum_{r=0}^n A_r\ihprav^r\over 1+ \sum_{r=1}^m B_r\ihprav^r} ,
\label{eq:Pade_def}
\end{equation}
whose Maclaurin series expansion matches that of $f(\ihprav)$ up to order $n+m$. This condition determines the coefficients $A_r$ and $B_r$~\cite{baker_theory_1964}.

The solid line in Fig.~\ref{fig:figEBC} represents the best Padé approximant, of order $[14/14]$, to the expansion~(\ref{ebc08.06.22-3}), truncated at $O(\ihprav^{30})$ and represented by the dashed line. Additionally, the red crosses and open blue circles depict numerical values of $P^{\rm SRM}_{\lambda(j)}$ for a few choices  of $j$ (equivalently $\ihprav$) and  for $N=500$ and $N=4600$ respectively. The figure readily shows the significant improvement provided by the Padé approximant near $\ihprav=1$ (see Appendix~\ref{app:semianalytic} for more details).

By replacing the integrand in Eq.~(\ref{eq: int P_lambda}) with its Padé approximant and performing the integral, we obtain the value $p_0(2) \approx 0.6499$. This leads us to conclude that the success probability for qubits approaches a finite value as $N\to\infty$, given~by
\begin{equation}
    \label{eq:numerics-qubit}
    P^{\rm SRM}_s\approx 0.6499+O(1/N).
\end{equation}

Using the same procedure, one can compute $p_0(d)$ for any value of the local dimension $d$. Results, along with their estimated error margins, for some selected values can be found in the second column of Table~\ref{ebc11.06.22-t2}.
\begin{table}[ht]
\begin{center}
\renewcommand{\arraystretch}{1.25}
\begin{tabular}{|c|| c| c |} 
 \hline
  $\ d \ $& $p_0(d)$ & $p^{\rm known}_0(d)$  \\ [0.5ex] 
 \hline\hline
 2 & $0.6499\pm 2\times 10^{-4}$ & $0.64991$ \\ [0.0ex]  
 \hline
3& $0.792308\pm 3\times 10^{-6}$ & $0.792311$\\ [0.0ex] 
 \hline
 4&$0.852860\pm 1\times 10^{-6}$&0.8528600 \\ [0.3ex] 
 \hline
 8&$0.9323011\pm 2\times 10^{-7}$&$0.9323011$ \\  [0.0ex]  
 \hline
\end{tabular}
\end{center}
\caption{Limiting success probability as $N\to\infty$ 
for the SRM protocol (second column) and average success probability assuming known domain states, computed from the QCP results in~\cite{sentis_quantum_2016} (third column).}
\label{ebc11.06.22-t2}
\end{table}

For the same values of $d$, the third column of Table~\ref{ebc11.06.22-t2} presents the averaged success probability assuming measurements can depend on the domain states.
Since our focus is solely on the limiting values as $N\to\infty$, we can rely on the asymptotic result provided in Ref.~\cite{sentis_quantum_2016}. In our current notation, this is expressed as
\begin{align}
p^{\rm known}_0(d)&=\!\!
\int {\rm d}\phi_0\,  {\rm d}\phi_1\, P_{\rm s}^{\rm QCP}(\phi_0,\phi_1)\nonumber \\[.5em]
&=\!\!\int_0^1 \dfrac{4(1\!-\!c^2)}{\pi^2} K^2(c^2)\, \mu_d(c^2)\, {\rm d}c^2,
\label{p_0 average}
\end{align}
where $c=|\braket{\phi_0}{\phi_1}|$ denotes the overlap of the states in each domain, $\mu_d(c^2)\,{\rm d}c^2$ represents the uniform (probability) measure for qudits~\cite{sentis_unsupervised_2019}, given by
\begin{equation}
\mu_d(c^2)=(d-1)(1-c^2)^{d-2},
\end{equation} 
and $K(x)$ is the complete elliptic function of the first kind~\cite{abramowitz_handbook_1968}.
The numerical integration has been carried out to high precision, ensuring that all digits shown in the third column of the table are significant.
As elucidated earlier, the agreement observed with the values in the second column tells us that they are optimal and  serves as a numerical verification of the asymptotic optimality of the SRM strategy.

After discussing the asymptotic regime of large $N$ and finite $d$, and to complete our analysis, we now turn our attention to the scenario of large local dimension $d$. This situation presents two distinct cases: one where $d\gg N\gg 1$ and another where $N\gg d\gg 1$. In both instances, the success probability is seen to approach unity (see~Appendix~\ref{app:semianalytic}), 
which is to be expected given that any two states randomly chosen from a highly dimensional Hilbert space are essentially orthogonal and therefore distinguishable.
However, computing the subleading term (of order $1/d$) in the former case is considerably more involved and does not provide significant insight. Therefore, we omit its presentation in this paper. 

Conversely, for the regime $N\gg d\gg 1$, this computation becomes straightforward, provided we extrapolate the agreement shown in Table~\ref{ebc11.06.22-t2} and conjecture that it will also hold in the limit $d\to\infty$. Then, the asymptotic expansion can be derived from Eq.~(\ref{p_0 average}) by observing that in this limit, $\mu_d(c^2)$ peaks near $c^2=0$, allowing us to expand the remainder of the integrand in powers of $c^2$. Integrating term by term yields
\begin{equation}
\lim_{N\to\infty}P_{\rm s}=1-\frac{1}{2d}+O(1/d^2).
\end{equation}

\section{Known state in one domain}
\label{sec:knownunknown}

In this section, we consider a situation where the state of the particles in one domain is known, specifically set to $|0\rangle$ without loss of generality, while the state of the particles in the other domain remains unknown. Consequently, measurements can depend solely on the known state $|0\rangle$. This setup is particularly relevant in the context of time series data. Consider a scenario where a particle source is designed to produce particles in the state $|0\rangle$. However, due to a malfunction, the source begins producing particles in an unspecified state after a certain period. Given a string of $N$ particles produced by this source, our objective is to develop a detection protocol that optimally identifies where along the string the state of the particles changed and determine its success probability.

In this scenario, the effective states of the string read [cf. Eqs.~\eqref{eq:twosymmetric}--\eqref{eq:dim-sym}]
\begin{equation}
    \label{eq:rho-k-u}
    \rho_k=\frac{\left(|0\rangle\langle 0|\right)^{\otimes N-k}\otimes \openone_{k}^{\mathrm{sym}}}{d_{k}^{\,\mathrm{sym}}}, 
\end{equation}
where, as in previous sections, we assume that the location of the edge, or state change, is uniformly distributed along the string. 
The specific~form~(\ref{eq:rho-k-u}) of the effective states implies several adjustments in the detection strategy developed in Section~\ref{sec:methodology}. For simplicity, we focus exclusively on qubits in this section, deferring details for arbitrary dimensions to Appendix \ref{app:k-u}.

We observe that the density matrices $\rho_k$ are diagonal in the basis $\{\,{|w^{n_1}_k}\rangle\,\}$ ($n_1=0,1,\ldots,k$; $k=1,2,\ldots,N$), which is defined as follows:
\begin{equation}
    \label{eq:basis-k-u}
   \ket{w^{n_1}_k}=\ket{0}^{\otimes{N-k}}\otimes\dfrac{\sum_\sigma P_\sigma\, |0\rangle^{\otimes n_0}\otimes|1\rangle^{\otimes n_1}}{\sqrt{\binom{k}{n_1}}}.
\end{equation}
Here, $n_0+n_1=k$, with $0\le n_0\leq k$. The set $\{P_\sigma\}$ stands for the unitary representation of $S_k$ on $({\mathbb C}^2)^{k}$ that permutes the $k$ particles and the sum extents only to those permutations $\sigma$ that do not leave the state
$ |0\rangle^{\otimes n_0}\otimes|1\rangle^{\otimes n_1}$ invariant. 
Clearly, only states with the same number $n_1$ may exhibit a non-vanishing overlap. This class of bases, known as Jordan bases~\cite{bergou_programmable_2006}, play a key role in programmable discriminators~\cite{sentis_multicopy_2010,he_programmable_2007}.

The optimal protocol, then, also involves two sequential measurements. First we measure the total number of excitations (number of states $|1\rangle$) to obtain  a value $n_1$, thereby projecting each state $\rho_k$ into the posterior state $|w_k^{n_1}\rangle$. Subsequently, a second measurement, akin to the corresponding step in the protocol for unknown states, is performed to discriminate among the states $|w^{n_1}_k\rangle$, each occurring with respective prior probabilities given by:
\begin{equation}
\eta^{n_1}_k=\frac{1}{N d^{\,\mathrm{sym}}_k}=\frac{1}{(k+1)N}.
\end{equation}

Following a similar approach as in Section~\ref{sec:structure}, we compute the probability of successful discrimination using the Gram matrices ${\sf G}^{n_1}$ of the sets $\{\ket{w^{n_1}_k}\}_{k=n_1}^{N}$, defined as the $n_1\times n_1$ Hermitian matrix with entries $G^{n_1}_{kk'}=\sqrt{\eta^{n_1}_k\eta^{n_1}_{k'}}\braket{w^{n_1}_k}{w^{n_1}_{k'}}$. From \eqref{eq:basis-k-u}, we readily find
\begin{equation}
    \label{eq:gram-norm-k-u}
    G^{n_1}_{k\,k'}=v^{n_1}_ku^{n_1}_{k'}:=\sqrt{\eta^{n_1}_k\eta^{n_1}_{k'} \dfrac{\binom{k}{n_1}}{\binom{k'}{n_1}}}, \quad k\le k' ,
\end{equation}
with $G^{n_1}_{k\,k'}=G^{n_1}_{k'\,k}$ if $k'<k$. 
The form of ${\sf G}^{n_1}$ indicates that they are semiseparable matrices, as observed with the Gram matrices in Section~\ref{sec:structure}. All the properties discussed there for ${\sf G}^\lambda$ also apply to ${\sf G}^{n_1}$. This parallelism between Eqs.~(\ref{eq:gram-norm-k-u}) and~(\ref{eq:def u v}) is noteworthy, where $n_1$ takes the role of $\lambda$, and only one binomial is present in Eq.~(\ref{eq:gram-norm-k-u}), reflecting that in this case only one domain of unknown states exists.

The calculations of the maximum and SRM success probabilities, the former obtained via SDP optimization, follow the methodology described earlier in the paper. Numerical results are depicted in Fig.~\ref{fig:all_data_ylf}, facilitating comparison with results obtained for unknown domain states. The blue circles represent the SDP results, calculated for $N$ up to 50, while the orange circles represent the numerical evaluation of the SRM success probability, extended to much larger values of $N$. 
Consistently, these points lie between those corresponding to unknown states in both domains (green and pink) and the asymptotic dashed line, which represents the opposite scenario where the states in both domains are known.
Remarkably, the blue and orange circles closely align, indicating that the SRM protocol remains nearly optimal in this scenario as well.

\section{Summary and conclusions}
\label{sec:5}

The detailed summary of our results is provided at the end of the introduction. Here, we complement it, offering a brief overview, and state our conclusions 

We introduced a task that we termed quantum edge detection, aimed at accurately locating quantum domain boundaries. As a first attempt to devise detection protocols, we focused on detecting boundaries between two quantum domains in a 1D system, where particles within each domain share the same pure state. 
Taking inspiration from recent advancements in quantum learning, we devised a protocol that is universal, meaning it does not rely on the specific states of the domains, and optimal, ensuring the highest success probability. Furthermore, we formulated a similar optimal protocol tailored for scenarios where the states in one domain are known, resulting in even higher success rates. For both protocols, this success rate converges to the same finite value as the string length approaches infinity. This limiting value, which can be derived from previous work on the quantum change point, requires knowing the state of each domain, suggesting that our detection protocols effectively learn the specifications of the domain states.

From a technical perspective, our findings confirm the asymptotic optimality of the square root measurement  
for quantum edge detection.
Moreover, our numerical analysis reveals that even with a small number of particles, this measurement closely approaches optimality. 
These conclusions are noteworthy because there are no known results regarding the optimality of the square root measurement for the type of discrimination problem encountered here.
Additionally, our results showcase the effectiveness of our technique, which uses weak Schur sampling, the Gram matrix formulation of discrimination, reducing the complexity of the optimization, and acceleration of perturbative expansions through Padé analysis. The latter, a technique not commonly employed in quantum information, proved highly valuable in deriving asymptotic limits for the problem at hand.

In conclusion, despite the inherent fragility of quantum data and the probabilistic nature of quantum measurements, which only provide limited partial information about a system's state, our study demonstrates precise edge detection in 1D is achievable. Specifically, we have shown that for large systems of qubits, exact edge detection can be achieved with a success rate of approximately 65\%. For qudits, this rate can reach 100\% as the local dimension approaches infinity. These results pave the way for exploring more physically relevant scenarios, such as edge detection in 2D and 3D, which could be applied to uncovering the spatial structure of complex systems. The protocols presented in this paper, albeit simple, hold promise for detecting malfunctions in quantum data sources.

Open problems along these lines include exploring other topologies (e.g., circle or torus in higher dimensions); devising efficient detection protocols based on local measurements; studying these problems under less demanding figures of merit, more suitable to quantum edge detection, such as Hamming distance or Wasserstein-type distances~\cite{villani_topics_2021}; and generalizing edge detection to quantum channels, which could be useful, for example, in finding domains of local magnetic fields affecting the orientation~of spins in a lattice.

Another important question that remains to be addressed in depth concerns the robustness of our protocol against imperfections in and along the string of particles. These may arise in various ways, e.g., the states may not be perfectly pure or may be slightly different along the string within each domain. A detailed analysis of this problem requires additional techniques beyond those introduced in this work and is currently under investigation. However, under reasonable assumptions~\cite{sentis_2012_quantum,sentis_2012_robust}, the domains' states may be thought of as if they undergo a small depolarization, in which case deviations from our results would be proportional to the depolarizing parameter, supporting the conclusion that our edge detection protocol, as it stands, is robust.

\section{Acknowledgements}

This work has been financially supported by  Ministerio de Ciencia e Innovación of the Spanish Goverment with funding from European Union Next GenerationEU (PRTR-C17.l1) and by the Generalitat of Catalunya,
by the Ministry of Economic Affairs and Digital Transformation of the Spanish Government through the QUANTUM ENIA project: Quantum Spain, by the European Union through the Recovery, Transformation and Resilience Plan - NextGenerationEU within the framework of the ``Digital Spain 2026 Agenda”, 
and by the Agencia Estatal de Investigación
MCIN/AEI/10.13039/501100011033 with projects PID2019-
107609GB-I00 and PID2022-141283NB-I00 with the support of FEDER funds. 
J.~C. also acknowledges support from the ICREA Academia award.
W.~G. acknowledges financial support from the Faculty of Sciences of the University of Los Andes during his stay at the UAB. Additionally, he extends his gratitude to the Quantum Information Group at the UAB for their hospitality.
\bibliography{EdgeDetection}

\onecolumn

\clearpage

\appendix
\twocolumn


\section{Schur basis and Gram matrices} \label{app:sixyeys}

In this appendix, we present a procedure for computing the states $|\Omega_k^\lambda\rangle$ and evaluating their overlaps. Our primary focus is on ordered sequences [as detailed in Eq.(\ref{eq:ordered sequence})], where $|\phi_0\rangle$ precedes $|\phi_1\rangle$. However, we extend our analysis to sequences lacking a definite order. We adhere to the notation and conventions of the main text whenever possible to maintain coherence.

Given $N-k$ pure states $|\phi_0\rangle$ and $k$ states $|\phi_1\rangle$, the ordered sequence remains invariant under any permutation belonging to the intransitive maximal subgroup $S_{N-k} \times S_k$. This group's action permutes the first $N-k$ elements (states) among themselves and the last $k$ elements among themselves without intermixing the two subsequences. This invariance will be used to prove that the Gram matrices ${\sf G}^\lambda$ are semi-separable.

In the Schur basis, a general state formed by the tensor product of such a sequence of states, regardless of whether it is ordered or disordered, is expressed as
\begin{equation}
\label{eq:V(ph0,ph1)}
\ket{\Phi_\s}=\sum_{\lambda} \ket{V_{\lambda}(\phi_0,\phi_1)} \otimes |\Omega_\s^{\lambda}\rangle,
\end{equation}
Here, instead of using $k$ as in Eq.~(\ref{eq:ordered sequence}), we label the states with a list of $0$s and $1$s, denoted by~$\s$, which specifies the positions of $|\phi_0\rangle$ and $|\phi_1\rangle$ in the tensor product. The state $\ket{V_\lambda (\phi_0,\phi_1)}$ belongs to the irreducible subspace $\lambda$ of ${\rm SU}(d)$, and its specific form depends on $|\phi_0\rangle$ and $|\phi_1\rangle$. In contrast, the state $|\Omega_\s^{\lambda}\rangle$ depends solely on the symmetries of the sequence $\s$ under permutations. 
Hence, to compute $|\Omega^\lambda_{\alpha}\rangle$, we can substitute $|\Phi_\s\rangle$ with a more convenient choice with the same permutational symmetries, simplifying the computation.  
One such option is the computational basis element $|\s\rangle$. Therefore, instead of Eq.~(\ref{eq:V(ph0,ph1)}), we~have:
\begin{equation}
      \ket{\s}= \sum_{\lambda} \ket{W_{\lambda}}\otimes|\Omega_{\s}^{\lambda}\rangle,
      \label{s = W Omega}
\end{equation}
where the expression for $|W_{\lambda}\rangle$ is inconsequential for this calculation. An important observation is that the state $|\Omega_{\s}^{\lambda}\rangle$, in the irrep subspace $\lambda$ of~$S_N$, is independent of the local dimension $d$. This, combined with the fact that we are only dealing with two-row YDs, allows us to further simplify the calculation by setting $d=2$. 

The elements of the Schur basis are conveniently denoted by $\ket{\lambda,w,q}$, where $w$  ($q$) labels the basis elements in the vector space of the irrep $\lambda$ of $\text{SU}(d)$~($S_N$). We refer to $w$ as the weight, which is defined as the total number of $1$s in the Semi-Standard Young Tableaux (SSYT)\footnote{\,SSYT are obtained by filling a YD with integer from $0$ to $d-1$ so that numbers do not decrease from left to right and strictly increase from top to bottom. In our case, we restrict the integer to take values $0$ and $1$ only.} 
containing only $0$s and $1$s associated with the irrep $\lambda$. %
For~${\rm SU}(2)$~the~weight is related~to the magnetic quantum number $m$ through the equation $m={N/2}-w$ (\mbox{recall} that in the main text we noted that \mbox{$j=N/2-\lambda$}). Similarly, each digit $\s_l$ in the binary sequence \mbox{$\s=(\s_1\s_2\dots \s_N)$} corresponds to the $z$-component of the spin through $s_l=1/2-\s_l$. Therefore, unless $w$ also coincides with the total number of $1$s in the sequence $\alpha$, the state $|\alpha\rangle$ is orthogonal to~$|\lambda,w,q\rangle$.

The last label, $q$, in $|\lambda,w,q\rangle$ specifies the basis elements of the irrep $\lambda$ of $S_n$. It can be identified with the Yamanouchi symbols of the various Standard Young Tableaux (SYT)\footnote{\,SYT are obtained by filling a YD with the integer $1,2,\dots,N$ so that numbers strictly increase from left to right and from top to bottom.}
 compatible with the YD denoted by $\lambda$ (i.e., by the partition $[N-\lambda,\lambda]$). 
 It consist of a sequence $q=(q_1q_2\dots q_N)$, specifying which row $q_l=1,2$ of the SYT contains the number $l$. For ${\rm SU}(2)$, $q_l=1$ ($q_l=2$) indicates that the coupling of the $l$-th spin to subsystem consisting of the first $l-1$ spins in the sequence increases (decreases) their total angular momentum by~$1/2$.
 
The Schur transformation, which represents the change of basis from $\{|\s\rangle\}$ to $\{|\lambda,w,q\rangle\}$ and reveals the structure shown in Eq.~(\ref{s = W Omega}), is given explicitly by:
\begin{equation}
    \ket{\s}=\sum_{\lambda,q} \ket{\lambda,w(\s),q} \braket{\lambda,w(\s),q}{\s}.
    \label{eq:chbas1}
\end{equation}
Here, we emphasize that the weight is determined by  $\s$, more precisely by $w(\s)=\sum_{l=1}^N \s_l$.
The matrix elements $\braket{\lambda,w(\s),q}{\s}$  can be computed recursively for any sequence $\s$ following~\cite{botero_universal_2018}. The recursive procedure involves `coupling' a qudit in a state of the computational basis ($|0\rangle$ and $|1\rangle$ for two row YD; as is our case) to a system of $n-1$ ($n=N,N-1,\dots,2$) qudits already in a state of the Schur basis. At step $n$, we have
\begin{equation}
    \ket{\lambda,w,q}=\sum_{\s_n=0}^1 \Gamma_{q_n,\s_n,n}^{\lambda,w} \ket{\lambda',w',q'}\otimes\ket{\s_n},
    \label{eq:cg1}
\end{equation}
where $\lambda'=\lambda-q_n+1$,  $w'=w-\s_n$, $q'$ is the sequence of length $n-1$ obtained by dropping $q_n$ from $q$, and $\Gamma_{q_n,\s_n,n}^{\lambda,w}$ is a Clebsch-Gordan coefficient. 
In~the case under consideration, it can be obtained straightforwardly from their well-known expressions for ${\rm SU}(2)$ and are collected in Table~\ref{tab:CGcoef}.
\begin{table}[h!]
    \centering
    \begin{tabular}{|c||c|c|}\hline
     \diagbox{$q_n$}{$\s_n$} & $0$ &$1$  \\ \hline\hline
     \rule{0pt}{15pt}$1$   & $\sqrt{\frac{n-\lambda-w}{n-2\lambda}}$ & $\sqrt{\frac{w-\lambda}{n-2\lambda}}$  \\ [1ex]  \hline
     \rule{0pt}{15pt}$2$ &  $-\sqrt{\frac{w-\lambda+1}{n-2\lambda+2}}$ & $\sqrt{\frac{n-\lambda-w+1}{n-2\lambda+2}}$ \\ [1ex] \hline
    \end{tabular}
    \caption{Clebsch-Gordan coefficients $\Gamma^{\lambda,w}_{q_n,\s_n,n}$.}
    \label{tab:CGcoef}
\end{table}

Noticing that $\langle\lambda,w,q|\alpha\rangle=\langle\alpha|\lambda,w,q\rangle$ (as they are real coefficients), we can express the state $|\s\rangle$ as:
\begin{equation}
\ket{\s}=
\sum_{\lambda,q} \prod_{n=1}^{N}\Gamma^{\lambda^{(n)},w^{(n)}}_{q_n,\s_n,n} \ket{\lambda,w(\s),q},
\label{eq:recursion}
\end{equation}
where $\lambda^{(n)}$ and $w^{(n)}$ denote respectively the partition (i.e., the irrep) and the weight at step $n$. From here, we can identify the expression of the state~$|\Omega^\lambda_\alpha\rangle$ in the $S_N$ basis $\{|\lambda,q\rangle\}_q$ alone, which is given by:
\begin{equation}
|\Omega^{\lambda}_{\s}\rangle= C_{\s}^{\lambda} \sum_{q} \prod_{n=1}^{N}\Gamma^{\lambda^{(n)},w^{(n)}}_{q_n,\s_n,n} \ket{\lambda,q}.
\label{eq:recursion2}
\end{equation}
Here, $C^\lambda_{\s}$ is a normalization constant that depends on the sequence $\alpha$ only through its weight $w=w(\alpha)$. It can be computed to be 
\begin{equation}
    C_{\s}^{\lambda}= \sqrt{\frac{\lambda!(1+N-\lambda)!}{(N-w)!(1+N-2\lambda)! w!}}\,.
    \label{eq: the C}
\end{equation}

As mentioned, Eqs.~(\ref{eq:recursion2}) and~(\ref{eq: the C}) apply to general sequences $\alpha$, whether ordered or disordered. Hereafter, we will exclusively focus on ordered sequences, as they are the only ones relevant for edge detection. 
So we go back to our original notation, $|\Omega^\lambda_k\rangle=|\Omega^\lambda_{\alpha(k)}\rangle$, where $\alpha(k)=(0^{N-k}1^k)$.
As discussed in the main text, our interest lies in obtaining the Gram matrix for the set  $\{|\Omega^\lambda_k\rangle\}_{k\in K_\lambda}$. This simply involves computing the overlaps $\langle\Omega^\lambda_k|\Omega^\lambda_{k'}\rangle$, rather than explicitly determining the states. These overlaps can be efficiently computed as follows. 

Consider the expression
\begin{equation}
    \braket{\Omega_{k}^{\lambda}}{\Omega_{k'}^{\lambda}}  \braket{\Omega_{k'}^{\lambda}}{\Omega_{k''}^{\lambda}}=
      \langle\Omega_{k}^{\lambda}|\,\Omega_{k'}^{\lambda}\,  |\Omega_{k''}^{\lambda}\rangle ,
    \label{eq:2innerprod 1}
\end{equation}
with $k\le k'\le k''$.
The operator $\Omega_{k'}^{\lambda}$ 
projects onto the invariant subspace of the intransitive maximal subgroup $S_{N-k'}\times S_{k'}$, discussed earlier in this appendix.  Thus, it is expressed as the product of projectors onto the invariant subspaces of $S_{N-k'}$ and~$S_{k'}$. 
Because $k\le k'$, $\langle\Omega_{k}^{\lambda} |$ lies on the invariant subspace of $S_{N-k'}$, and because $k'\le k''$, $|\Omega_{k''}^{\lambda}\rangle$ belongs to the invariant subspace of $S_{k'}$. Therefore, $\Omega_{k'}^{\lambda}$ acts as the identity in Eq.~(\ref{eq:2innerprod 1}), and we obtain 
\begin{equation}
 \braket{\Omega_{k}^{\lambda}}{\Omega_{k'}^{\lambda}}  \braket{\Omega_{k'}^{\lambda}}{\Omega_{k''}^{\lambda}}=\braket{\Omega_{k}^{\lambda}}{\Omega_{k''}^{\lambda}},
\end{equation}
provided $k\le k'\le k''$.
Setting $k$ to its smallest possible value, $k=\lambda$, we derive the relation
\begin{equation}
  \braket{\Omega_{k'}^{\lambda}}{\Omega_{k''}^{\lambda}} = {  \braket{\Omega_{\lambda}^{\lambda}}{\Omega_{k''}^{\lambda}} \over  \braket{\Omega_{\lambda}^{\lambda}}{\Omega_{k'}^{\lambda}}  }.
    \label{eq:2innerprod}
\end{equation}
This proves semi-separability of ${\sf G}^\lambda$, since it implies [cf. Eqs.~(\ref{eq:overlap-u-u}) and~(\ref{eq:def u v})]
\begin{equation}
u_k={\sqrt\eta^\lambda_k\over \langle\Omega^\lambda_\lambda|\Omega^\lambda_k\rangle},
\quad
v_k=\sqrt\eta^\lambda_k\, \langle\Omega^\lambda_\lambda|\Omega^\lambda_k\rangle .
\label{eq:u and v gen}
\end{equation}

Equation (\ref{eq:u and v gen}) simplifies the calculation of the Gram matrices by reducing the required number of overlaps to $|K_\lambda|$. Moreover, the state $|\Omega^\lambda_\lambda\rangle$ is easily computed, having the simple expression:
\begin{equation}
    |\Omega_{\lambda}^{\lambda}\rangle= (-1)^{\lambda}\, |\lambda,(1^{N-\lambda}2^\lambda)\rangle.
    \label{eq:vector1}
\end{equation}
This state is an element of the Schur basis (with a sign change for odd partitions), and therefore its inner product with any vector $|\Omega_k^\lambda\rangle$ is simply $(-1)^{\lambda}$ times the coefficient corresponding to \mbox{$q=(1^{N-\lambda}2^\lambda)$} in Eq.~\eqref{eq:recursion2}. Explicitly:
\begin{align}
\kern-.4em\nonumber\braket{\Omega_{k}^{\lambda}}{\Omega_{\lambda}^{\lambda}}&= (-1)^{\lambda}C^{\lambda}_{\alpha(k)} \left(\prod_{n=1}^{N-k} \Gamma^{0,0}_{1,0,n}\!\right)\!  \times \\[.5em]
\nonumber 
&\kern 3.25em \left( \prod_{n=N-k+1}^{N-\lambda}\kern-1em \Gamma^{0,n-N+k}_{1,1,n}\!\right)\!\times\\[.5em]
\nonumber 
&\kern.75em \left(\prod_{n=N-\lambda+1}^{N}\kern-1em \Gamma^{n-N+\lambda,n-N+k}_{2,1,n}\!\right)
\\[.5em]
&=\sqrt{\frac{\binom{N-k}{\lambda}}{\binom{N-\lambda}{\lambda}\binom{k}{\lambda}}}.
\label{eq:overlaplambda}
 \end{align}
Substituting this result in Eq.~(\ref{eq:u and v gen}) we obtain the Gram matrix ${\sf G}^\lambda$. Likewise, the overlaps can be computed from Eq.~\eqref{eq:2innerprod}, yielding 
\begin{equation}
\braket{\Omega_{k}^{\lambda}}{\Omega_{k'}^{\lambda}}=\sqrt{\frac{\binom{k}{\lambda}\binom{N-k'}{\lambda}}{\binom{k'}{\lambda}\binom{N-k}{\lambda}}},
\quad k\le k'.
\end{equation}
This expression agrees with that of the overlaps of two different couplings of three angular momenta~\cite{edmonds_angular_1996}, using the Wigner \mbox{$6j$-symbols}.



\section{Asymptotic results}
\label{app:semianalytic}

The method we employed to calculate the asymptotic success probability was previously discussed in Section~\ref{sec:optimal}. Here, we present some missing steps, including intermediate expressions (many of which are organized in tables), to assist interested readers in replicating our findings. We use the variables $j = N/2 - \lambda$ and $m = N/2 - w$ thoughout the appendix. While denoting angular momentum and magnetic quantum number, respectively, only for $d=2$, the use of $j$ and $m$ in deriving the asymptotic success probability has proven notably advantageous, resulting in simpler and more transparent expressions also for $d>2$.

The entries of the Gram matrix, ${\sf G} := {\sf G}^{\lambda}$, where we drop the label $\lambda$ to simplify the notation, are given in Eqs.~(\ref{eq:overlap-u-u}) and~(\ref{eq:def u v}). They can be expressed as follows:
\begin{multline}
G_{ k\, k'}
={2j+1\over {N\over2}+j+1}\times \\
 {{d+{N\over2}-j-2\choose d-2} {d+{N\over2}+j-1\choose d-1}\over
\sqrt{
{d+ k-1\choose d-1}{d+\bar k-1\choose d-1}
{d+ k'-1\choose d-1}{d+\bar k'-1\choose d-1}
}
}\times
\\
\sqrt{{k\choose{N\over2}-j}{\bar k'\choose{N\over2}-j}\over{k'\choose{N\over2}-j}{\bar k\choose{N\over2}-j}},
\label{ebc06.06.22-1}
\end{multline}
where the last ($d$-independent) factor can be clearly identified with the overlap $\langle\Omega^\lambda_k|\Omega^\lambda_{k'}\rangle$, while the remaining ($d$-dependent) factors correspond to the priors $\eta^\lambda_k$ and $\eta^\lambda_{k'}$.
Here $\bar k = N-k$, $\bar k' = N-k'$, and the range of $k$ and $k'$ is $(N/2)-j \le k \le (N/2)+j$. We assume $k \le k'$ ($\bar k' \le \bar k$); for $k' \le k$, we simply exchange $k$ with $k'$ and $\bar k$ with $\bar k'$, i.e., ${\sf G}$ is symmetric. The Gram matrix is also persymmetric, meaning it is symmetric with respect to the anti-diagonal. This implies that ${\sf G}$ is invariant under the exchange $k, k' \leftrightarrow \bar k, \bar k'$.

The scaled Gram matrix, used to derive the asymptotic success probability,  is [cf.~Eq.~(\ref{eq:rescaling})]:
\begin{equation}
\tilde {\sf G}={(N/2)^2\over (d-1)(2j+1)} {\sf G} .
\end{equation}
The inverse of $\tilde{\sf G}$ is tridiagonal (because $\tilde{\sf G}$ is semiseparable). Its diagonal and super-diagonal entries are respectively given by the sequences:
\begin{equation}
[\tilde {\sf G}^{-1}]_{\rm diag}\!=\!
{\mathscr B}^j_m\frac{j(j\!+\!1)\!+\!\frac{N}{2}\!\left(\!\frac{N}{2}\!+\!1\!\right)\!-\!2 m^2}{\left(\!\frac{N}{2}\!-\!j\!\right)\left(\!\frac{N}{2}\!+\!j\!+\!1\!\right)}
\end{equation}
($-j\le m\le j$),
and
\begin{multline}
[\tilde {\sf G}^{-1}]_{\text{super}}\!=\! -\sqrt{{\mathscr B}^j_m {\mathscr B}^j_{m+1}}  \times  \\[.5em]
\frac{\sqrt{\!\left(\!\frac{N}{2}\!-\!m\!\right)\!\left(\!\frac{N}{2}\!+\!m\!+\!1\!\right)\!(j\!-\!m)(j\!+\!m\!+\!1)}}{\left(\!\frac{N}{2}\!-\!j\!\right)\!\left(\!\frac{N}{2}\!+\!j\!+\!1\!\right)}
\end{multline}
($-j\le m\le j-1$), where
\begin{multline}
{\mathscr B}^j_m=\frac{1\!+\!\frac{N}{2}\!+\!j}{\left(N/2\right)^2}\times \\
\kern-1em\frac{\left(\!\frac{N}{2}\!-\!j\!\right)!\left(\!\frac{N}{2}\!+\!j\!\right)!}{\left(\!\frac{N}{2}\!-\!j\!+\!d\!-\!2\!\right)!\left(\!\frac{N}{2}\!+\!j\!+\!d\!-1\!\right)!} \times
\\
\frac{\left(\!\frac{N}{2}\!-\!m\!+\!d\!-\!1\!\right)!\left(\!\frac{N}{2}\!+\!m\!+\!d\!-\!1\!\right)!}{\left(\!\frac{N}{2}\!-\!m\!\right)!\left(\!\frac{N}{2}\!+\!m\!\right)!}.
\end{multline}

As explained in Section~\ref{sec:optimal}, we introduce a scaled version of $j$, defined as $\ihprav = j/(N/2)$, and compute the Maclaurin series expansion of the joint probability $P(x) := P^{\rm SRM}_{\lambda(x)}$ (indices are dropped to avoid clutter) up to a high order, $2R$, in~$x$. At the leading order in $1/N$, the joint probability is given by the series: %
\begin{equation}
    \frac{N}{2}P(\ihprav)= \sum_{r=1}^\infty a_r \ihprav^{2r}.
    \label{eq:series app}
\end{equation}
The (exact) coefficients $a_r$ have been computed up to $r=15$ or $r=18$, depending on $d$, using {\em Wolfram Mathematica 13.3}, with the corresponding code available in~\cite{github}. For $d=2,3,4$, and~$8$, they are collected in Tables~\ref{tab:appcoeff2}--\ref{tab:appcoeff8}.

\begin{table}[ht]
\begin{center}
\renewcommand{\arraystretch}{1.2}
\begin{tabular}{|c|| c| c |} 
 \hline
  $\ l \ $& $a_{2l-1}$ & $a_{2l}$  \\ [0.5ex] 
 \hline
$1$&$2$&$0$\\[0.3ex]\hline$2$&$-\frac{1}{30}$&$-\frac{1}{35}$\\[0.3ex]\hline$3$&$-\frac{229}{10080}$&$-\frac{101}{5544}$\\[0.3ex]\hline$4$&$-\frac{5725}{384384}$&$-\frac{5111}{411840}$\\[0.3ex]\hline$5$&$-\frac{554293}{52715520}$&$-\frac{2137221}{236487680}$\\[0.3ex]\hline$6$&$-\frac{249919231}{31783944192}$&$-\frac{76576105}{11076222976}$\\[0.3ex]\hline$7$&$-\frac{10870862389}{1772195676160}$&$-\frac{8413125001}{1533630873600}$\\[0.3ex]\hline$8$&$-\frac{1506595197973}{304973453721600}$&$-$
\\[0.3ex]
 \hline
\end{tabular}
\end{center}
\caption{Coefficients $a_r$ of the MacLaurin series of $(N/2)P(\ihprav)$ defined in Eq.~\eqref{eq:series app} for $d=2$.}
\label{tab:appcoeff2}
\end{table}

\begin{table}[ht]
\begin{center}
\renewcommand{\arraystretch}{1.2}
\begin{tabular}{|c|| c| c |} 
 \hline
  $\ l \ $& $a_{2l-1}$ & $a_{2l}$  \\ [0.5ex] 
 \hline
$1$&$4$&$-\frac{8}{3}$\\[0.3ex]\hline$2$&$-\frac{1}{15}$&$0$\\[0.3ex]\hline$3$&$\frac{3}{560}$&$\frac{3}{616}$\\[0.3ex]\hline$4$&$\frac{739}{192192}$&$\frac{307}{102960}$\\[0.3ex]\hline$5$&$\frac{1044347}{448081920}$&$\frac{874097}{472975360}$\\[0.3ex]\hline$6$&$\frac{23645057}{15891972096}$&$\frac{5047729}{4153583616}$\\[0.3ex]\hline$7$&$\frac{891227339}{886097838080}$&$\frac{168502693}{200038809600}$\\[0.3ex]\hline$8$&$\frac{3749131111}{5258162995200}$&$\frac{9600936701}{15756961775616}$\\[0.3ex]\hline$9$&$\frac{1398178715421307}{2662296261608079360}$&$-$
\\[0.3ex]
 \hline
\end{tabular}
\end{center}
\caption{Coefficients $a_r$ of the MacLaurin series of $(N/2)P(\ihprav)$ defined in Eq.~\eqref{eq:series app} for $d=3$.}
\label{tab:appcoeff3}
\end{table}

\begin{table}[ht]
\begin{center}
\renewcommand{\arraystretch}{1.2}
\begin{tabular}{|c|| c| c |} 
 \hline
  $\ l \ $& $a_{2l-1}$ & $a_{2l}$  \\ [0.5ex] 
 \hline
$1$&$6$&$-8$\\[0.3ex]\hline$2$&$\frac{31}{10}$&$\frac{3}{35}$\\[0.3ex]\hline$3$&$\frac{9}{1120}$&$0$\\[0.3ex]\hline$4$&$-\frac{125}{128128}$&$-\frac{25}{27456}$\\[0.3ex]\hline$5$&$-\frac{3875}{5431296}$&$-\frac{51075}{94595072}$\\[0.3ex]\hline$6$&$-\frac{393277}{963149824}$&$-\frac{3454141}{11076222976}$\\[0.3ex]\hline$7$&$-\frac{125777259}{521234022400}$&$-\frac{145143689}{766815436800}$\\[0.3ex]\hline$8$&$-\frac{15293179829}{101657817907200}$&$-\frac{249322259}{2059733565440}$\\[0.3ex]\hline$9$&$-\frac{174838753670533}{1774864174405386240}$&$-$
\\[0.3ex]
 \hline
\end{tabular}
\end{center}
\caption{Coefficients $a_r$ of the MacLaurin series of $(N/2)P(\ihprav)$ defined in Eq.~\eqref{eq:series app} for $d=4$.}
\label{tab:appcoeff4}
\end{table}

\begin{table}[ht]
\begin{center}
\renewcommand{\arraystretch}{1.2}
\begin{tabular}{|c|| c| c |} 
 \hline
  $\ l \ $& $a_{2l-1}$ & $a_{2l}$  \\ [0.5ex] 
 \hline
$1$&$14$&$-56$\\[0.3ex]\hline$2$&$\frac{3353}{30}$&$-127$\\[0.3ex]\hline$3$&$\frac{120347}{1440}$&$-\frac{11675}{396}$\\[0.3ex]\hline$4$&$\frac{226243}{54912}$&$\frac{35777}{411840}$\\[0.3ex]\hline$5$&$\frac{7310429}{896163840}$&$\frac{119175}{94595072}$\\[0.3ex]\hline$6$&$\frac{1121953}{4540563456}$&$\frac{83349}{1582317568}$\\[0.3ex]\hline$7$&$\frac{12326391}{1265854054400}$&$0$\\[0.3ex]\hline$8$&$-\frac{20469449}{11295313100800}$&$-\frac{61408347}{35015470612480}$\\[0.3ex]\hline$9$&$-\frac{24207914731}{17927920953589760}$&$-$
\\[0.3ex]
 \hline
\end{tabular}
\end{center}
\caption{Coefficients $a_r$ of the Maclaurin series of $(N/2)P(\ihprav)$ defined in Eq.~\eqref{eq:series app} for $d=8$.}
\label{tab:appcoeff8}
\end{table}

The Padé approximants (referred to as Padés hereafter) to $(N/2)P(x)$ can be computed from 
those tables, as outlined in Section~\ref{sec:optimal}. Once again, we relied on {\em Mathematica} to perform this tedious calculation. 
Diagonal Padés, of the form shown in Eq.~(\ref{eq:Pade_def}) with $n=m$, have proven to deliver lower uncertainly and better stability for the problem at hand. We express them as:
\begin{equation}
{\mathscr P}_{2s}(x):={ \sum_{r=0}^{s} A_r x^{2r}\over 1+ \sum_{r=1}^{s} B_r x^{2r}} \,.
\end{equation}
%

\begin{table}[ht]
\begin{center}
\renewcommand{\arraystretch}{1.2}
\begin{tabular}{|c|| c| c |} 
 \hline
  $\ r \ $& $A_{r}$ & $B_{r}$  \\ [0.0ex] 
 \hline
$1$&$2.$&$-3.47961$\\[0.3ex]\hline$2$&$-6.95921$&$4.68403$\\[0.3ex]\hline$3$&$9.33473$&$-3.03572$\\[0.3ex]\hline$4$&$-5.98403$&$0.951594$\\[0.3ex]\hline$5$&$1.82375$&$-0.125282$\\[0.3ex]\hline$6$&$-0.22237$&$0.00507854$\\[0.3ex]\hline$7$&$0.00725633$&$-0.0000178728$
\\  [0.0ex]  
 \hline
\end{tabular}
\end{center}
\caption{Coefficients of the Pad\'e approximant ${\mathscr P}_{14}(\ihprav)$ for $d=2$.}
\label{tab:padecoeff2}
\end{table}

\begin{table}[ht]
\begin{center}
\renewcommand{\arraystretch}{1.2}
\begin{tabular}{|c|| c| c |} 
 \hline
  $\ r \ $& $A_{r}$ & $B_{r}$  \\ [0.0ex] 
 \hline
$1$&$4.$&$-1.56531$\\[0.3ex]\hline$2$&$-8.9279$&$-0.406743$\\[0.3ex]\hline$3$&$2.48052$&$1.89668$\\[0.3ex]\hline$4$&$8.77573$&$-1.14072$\\[0.3ex]\hline$5$&$-9.58824$&$0.240021$\\[0.3ex]\hline$6$&$3.87205$&$-0.0158881$\\[0.3ex]\hline$7$&$-0.633517$&$0.000108498$\\[0.3ex]\hline$8$&$0.0319437$&$0.00000116197$
\\  [0.0ex]  
 \hline
\end{tabular}
\end{center}
\caption{Coefficients of the Pad\'e approximant ${\mathscr P}_{16}(\ihprav)$ for $d=3$.}
\label{tab:padecoeff3}
\end{table}

\begin{table}[ht]
\begin{center}
\renewcommand{\arraystretch}{1.2}
\begin{tabular}{|c|| c| c |} 
 \hline
  $\ r \ $& $A_{r}$ & $B_{r}$  \\ [0.0ex] 
 \hline
$1$&$6.$&$-2.7713$\\[0.3ex]\hline$2$&$-24.6278$&$2.85673$\\[0.3ex]\hline$3$&$42.4108$&$-1.33735$\\[0.3ex]\hline$4$&$-39.3833$&$0.276189$\\[0.3ex]\hline$5$&$20.9823$&$-0.0205167$\\[0.3ex]\hline$6$&$-6.2558$&$0.000174461$\\[0.3ex]\hline$7$&$0.928717$&$0.00000312601$\\[0.3ex]\hline$8$&$-0.0502589$&$0.0000000704169$
\\  [0.0ex]  
 \hline
\end{tabular}
\end{center}
\caption{Coefficients of the Pad\'e approximant ${\mathscr P}_{16}(\ihprav)$ for $d=4$.}
\label{tab:padecoeff4}
\end{table}

For each $d$, we have analyzed the behavior of the sequence ${\mathscr P}_{2}(\ihprav), {\mathscr P}_{4}(\ihprav), \dots$ that can be derived from Tables~\ref{tab:appcoeff2}--\ref{tab:appcoeff8} and concluded that it exhibits convergence and stability across the entire range \mbox{$0 \leq \ihprav \leq 1$}. From our analysis, we estimated the corresponding asymptotic success probability by integrating the highest-order Padé of the sequence, ${\mathscr P}_{\rm high}(\ihprav)$, as in Eq.~(\ref{eq: int P_lambda}):
\begin{equation}
P_{\rm s} \sim p_0(d) \approx \int_0^1 {\mathscr P}_{\rm high}(\ihprav) d\ihprav.
\end{equation}
These estimated values correspond to the lower margins provided in the second column of Table~\ref{ebc11.06.22-t2}.
For completeness, 
Tables~\ref{tab:padecoeff2}--\ref{tab:padecoeff4} gather the coefficients of the highest-order diagonal Padés
for $d=2,3,4$. However, for higher dimensionality, the Padés do not seem to offer any improvement over the Maclaurin series expansion, and they are omitted here.

Alternatively, one could find the primitive of Eq.~(\ref{eq:series app}), represented by the series
\begin{equation}
Q(x) = \sum_{r=1}^\infty \frac{a_r}{2r+1} \ihprav^{2r+1},
\end{equation}
and then, to estimate the asymptotic success probability \mbox{$P_{\rm s} \sim p_0(d)= Q(1)$}, apply Padé acceleration to the truncated series approximation obtainable from Tables~\ref{tab:appcoeff2}--\ref{tab:appcoeff8}. Note that $Q(\ihprav)$ is an odd function, preventing the construction of diagonal Padés. 
For this reason, we examine the behavior of the Padé sequence $\{{\mathscr Q}^{2n-1}_{2n}(\ihprav), {\mathscr Q}^{2n+1}_{2n}(\ihprav)\}_{n}$, where ${\mathscr Q}^{n}_{m}(\ihprav) = [n/m]_Q(\ihprav)$. This sequence also exhibits rapid convergence and stability in the unit interval and, consequently, we estimate the asymptotic success probability from its highest-order Padé, ${\mathscr Q}_{\rm high}$, as
\begin{equation}
P_{\rm s} \sim p_0(d) \approx {\mathscr Q}_{\rm high}(1).
\end{equation}
The results are the upper margins shown in the second column of~Table~\ref{ebc11.06.22-t2}. 

We estimate the accuracy of this method by computing the difference between the upper and lower margins obtained with the two described alternatives. In all cases this yields an estimated uncertainty of less than $0.03\%$, which we see as a clear indication of the consistency of our Padé analysis.

For a local dimension of around 20 or higher, certain Maclaurin coefficients exhibit significant growth, alternating signs, and the series becomes unstable. While Padé approximants offer some assistance in the analysis, the estimated accuracy diminishes, requiring a different approach to obtain precise results for the success probability.

For asymptotically large local dimensions,  assuming additionally that $d\gg N$, the Gram matrix becomes 
\begin{multline}
G_{ k\, k'}={(2j+1)\sqrt{\bar k!\bar k'!k!k'!}\over\left({N\over2}-j\right)!\left({N\over2}+j+1\right)!}\times \\[.5em]
\sqrt{{k\choose{N\over2}-j}{\bar k'\choose{N\over2}-j}\over{k'\choose{N\over2}-j}{\bar k\choose{N\over2}-j}}+O\left(1/d\right),
\label{ebc06.06.22-2}
\end{multline}
where $k\le k'$. 
One can follow a similar procedure as outlined to obtain an expansion analogous to Eq.~(\ref{eq:P lambda exp}). Its first terms are given by:
\begin{multline}
P_\lambda={2(2j\!+\!1)^2\over N^2}\left[1 -
{2(2j^2\!+\!2j\!+\!3)\over 3N}+\right.\\
\left.{4(j^2\!+\!j\!+\!3)(4j^2\!+\!4j\!+\!5)\over 15 N^2}+\dots\right] .
\end{multline}
A glimpse at the few terms presented in this equation already reveals that in this regime, the appropriate scaling of $j$ to determine the asymptotic behavior is $y=j/\sqrt{N/2}$. In this case, the Maclaurin series corresponding to the leading term in $1/N$ is:
\begin{align}
{N\over2}P(y)&=\sum_{l=1}^\infty {(-1)^{l+1}(2y^2)^l\over(2l-1)!!}\label{ebc03.06.22-3a}\\
&=2y F(y),
\label{ebc03.06.22-3b}
\end{align}
where $F(y)$ is identified as Dawson's integral, defined as~\cite{abramowitz_handbook_1968}:
\begin{equation}
F(y):={\rm e}^{-y^2}\int_0^y {\rm e}^{t^2}dt.
\label{ebc07.06.22-1}
\end{equation}
The asymptotic probability of success is computed as $P_{\rm s} \sim 2 \sqrt{N/2} \int_0^1 x F\big(\sqrt{N/2}\,x\big) dx$ (since $y=\sqrt{N/2}\, x$), which approaches unity as $N$ tends to infinity.


\section{Gram matrix for known state in one domain}
\label{app:k-u}
In this appendix, we derive the expression of the Gram matrices for a general local dimension $d$, assuming that all particles within one of the two domains share the same known state, which we take to be $|0\rangle$ without loss of generality.

In this context, Eq.~\eqref{eq:basis-k-u} assumes a more general form:
\begin{multline}
 |w^{\mathbf n}_k\rangle= { 1\over \sqrt{  \binom{k}{ {\mathbf n} }  }  }\;|0\rangle^{  N-k}    \\
\sum_{\sigma}\!P_\sigma|0\rangle^{n_0} |1\rangle^{n_1}\!\cdots\!|d\!-\!1\rangle^{n_{d-1}}  ,
\label{eq:w d}
\end{multline}
where we dropped the tensor-product symbols to shorten the expression.
In this equation, the count of particles in each of the states $|0\rangle, |1\rangle,\dots,|d-1\rangle$ is represented by a multi-index label ${\mathbf n}=(n_0,n_1,\ldots,n_{d-1})$, where \mbox{$\sum_{l=0}^{d-1} n_l=k$}. The operators $P_\sigma$, acting on $({\mathbb C}^d)^{\otimes k}$, \mbox{constitute} the unitary representation of the symmetric group $S_k$, which permutes the $k$ particles among themselves. The summation over $\sigma$ includes only permutations that do not act trivially on $|0\rangle^{n_0}\cdots\! |d\!-\!1\rangle^{n_{d-1}}$, and the multinomial coefficients are given by
\begin{equation}
{k \choose {\mathbf n}}:={ k! \over \prod_{l=0}^{d-1} n_l !}\, .
\end{equation}

It is useful (and meaningful) to assign a new label $\tilde{\mathbf{n}}$ to each label ${\mathbf{n}}$, according to
\begin{equation}
{\mathbf n}\to \tilde{\mathbf n}=(N-k+n_0,n_1,\ldots,n_{d-1}).
\end{equation}
The zeroth component, $\tilde n_0=N-k+n_0$, denotes the number of particles in the state $|0\rangle$ within a string of length $N$ that is in the state $|w^{\mathbf{n}}_k\rangle$. 
It is apparent that only states sharing the same entire label $\tilde{\mathbf{n}}$ may have a non-vanishing overlap. Therefore, after an initial measurement, projecting the state of the string onto a subspace with a definite $\tilde{\mathbf{n}}$, the conditional Gram matrices of the posterior states simply read:
\begin{align}
    G_{k\, k'}^{\,\tilde{\mathbf n}}&={\braket{w^{\mathbf n}_k}{w^{{\mathbf n}'}_{k'}}\over \sqrt{N d^{\,\rm sym}_k d^{\,\rm sym}_{k'}}}\nonumber\\
    &=\dfrac{1}{\sqrt{N d_{k}^\textrm{\,sym}d_{k'}^\textrm{\,sym}}}\sqrt{\dfrac{\binom{k}{\mathbf n}}{\binom{k'}{{\mathbf n}'}}}\, ,
    \label{eq:Gram_ku}
\end{align}
where  $d_k^{\mathrm{\,sym}}$ is given by Eq.~\eqref{eq:dim-sym} and it is assumed that $\tilde{\mathbf n}=\tilde{\mathbf n}'$ and $k\le k'$ (for $k'<k$ simply exchange $k$ and $k'$).
This expression can be further simplified by noting that, due to the condition $\tilde{\mathbf{n}}=\tilde{\mathbf{n}}'$, the ratio of multinomials reduces to a ratio of binomials,  ${\raisebox{-.1em}{\scriptsize $k$}\choose \raisebox{.1em}{\scriptsize $n_0$}}/{\raisebox{-.14em}{\scriptsize $k'$}\choose \raisebox{.14em}{\scriptsize $n'_0$}}$, showing that ${\sf G}^{\tilde{\mathbf n}}$ is  independent of $n_l$ for $l>0$.

This simplification enables us to aggregate identical contributions and label the matrices by the only relevant parameter $\tilde{n}_0$ (i.e., the number of particles in the string whose state is $|0\rangle$). Thus, we have:
\begin{align}
    \nonumber G_{k\, k'}^{\tilde{n}_0}&= %
  \!\! \! \sum_{n_1,n_2,\dots} \!\!\! G_{k\,k'}^{\tilde{{\mathbf n}}}\\
    &=\dfrac{\binom{N-\tilde{n}_0+d-2}{d-2}}{\sqrt{ N d_{k}^\textrm{\,sym}d_{k'}^\textrm{\,sym}}}\sqrt{\dfrac{\binom{k}{N-\tilde{n}_0}}{\binom{k'}{N-\tilde{n}_0}}}\, .
\end{align}
The expression $N-\tilde{n}_0$ counts the `excitations' in the string, referring to the number of particles in one of the states $|1\rangle$, $|2\rangle$, up to $|d-1\rangle$. For qubits, this number corresponds to $n_1$, thereby recovering Eq.~(\ref{eq:gram-norm-k-u}).
\onecolumn
\end{document}